# Optimizing an Utility Function for Exploration / Exploitation Trade-off in Context-aware Recommender System

Djallel Bouneffouf
Télécom SudParis
9, rue Charles Fourier
91011 Evry, France
Djallel.Bouneffouf@it-sudparis.eu

**Abstract**

In this paper, we develop a dynamic exploration/ exploitation (exr/exp) strategy for contextual recommender systems (CRS). Specifically, our methods can adaptively balance the two aspects of exr/exp by automatically learning the optimal tradeoff. This consists of optimizing a utility function represented by a linearized form of the probability distributions of the rewards of the clicked and the non-clicked documents already recommended. Within an offline simulation framework we apply our algorithms to a CRS and conduct an evaluation with real event log data. The experimental results and detailed analysis demonstrate that our algorithms outperform existing algorithms in terms of click-through-rate (CTR).

## 1. Introduction

Most professional mobile users acquire and maintain a large amount of content in their repository, for instance, for filtering news documents or for the display of advertisements.

Moreover, the content of such repository changes dynamically, undergoes frequent insertions and deletions.

In such a setting, it is crucial to identify interesting content for users. For instance, the system must promptly identify the importance of new documents, while also adapting to the fading value of existing, old documents.

Some works found in the literature (Wei et al., 2010; Lihong et al., 2010) address this problem as a need for balancing exr/exp studied in the "bandit algorithm". A bandit algorithm B exploits its past experience to select documents that appear more frequently.

Besides, these seemingly optimal documents may in fact be suboptimal, because of the imprecision in B's knowledge.

In order to avoid this undesired situation, B has to explore documents by actually choosing seemingly suboptimal documents so as to gather more information about them.

Exploitation can decrease short-term user's satisfaction since some suboptimal documents may be chosen.

However, obtaining information about the documents' average rewards (i.e., exploration) can refine B's estimate of the documents' rewards and in turn increase long-term user's satisfaction.

Clearly, neither a purely exploring nor a purely exploiting algorithm works best in general, and a good tradeoff is needed.

One classical solution to the multi-armed bandit problem is the ε-greedy strategy (Watkins, 1986). With probability 1-ε, this algorithm chooses the best action based on current knowledge; and with probability ε, it chooses any other action uniformly.

The parameter essentially controls the tradeoff between exploitation and exploration. One drawback of this algorithm is that the optimal value is difficult to decide in advance. Instead, we propose an algorithm that achieve this goal by adaptively balance this tradeoff.

In the this method, we extend the ε-greedy strategy with an update of the exr/exp-tradeoff by varying the exploration according to the optimization of the function used to evaluate the recommender systems performance (utility function), this strategy allows making an exploration when we get a high clicked documents' distribution and make exploitation when we get a low clicked-document.

In this paper, we make an offline simulation framework from real event logs in a CRS, and evaluate the impacts of exr/exp on the system's performance.

## 2. Related Work

We refer, in the following, recent recommendation techniques that tackle both issues, namely: making dynamic exr/exp and considering the user's situation on the exr/exp strategy.

### 2.1 Bandit Algorithms Overview (ε-greedy)

The ε-greedy is the most widely used strategy to solve the bandit problem and is first described by (Watkins, 1986).

The ε-greedy strategy consists of choosing a random lever with ε-frequency, and choosing otherwise the lever with the highest estimated mean, the estimation being based on the rewards observed thus far. ε must be in the open interval [0, 1] and its choice is left to the user.

Methods that imply a binary distinction between exploitation (the greedy choice) and exploration (uniform probability over a set of levers) are known as semi-uniform methods.

The simplest variant of the ε-greedy strategy is what (Even-Dar et al., 2003) and (Mannor & Tsitsiklis, 2003) refer to as the **ε-beginning strategy**.

The ε-beginning strategy consists in doing the exploration all at once at the beginning. For a given number $T \in N$ of rounds, the levers are randomly pulled during the εT first rounds (pure exploration phase).

During the remaining $(1-\varepsilon)T$ rounds, the lever of highest estimated mean is pulled (pure exploitation phase). Here too, ε must be in the open interval [0, 1] and its choice is left to the user.

A natural variant of the ε-greedy strategy is what Cesa-Bianchi and Fisher (Mannor & Tsitsiklis, 2003) call the **ε-decreasing strategy**.

The ε-decreasing strategy consists of using a decreasing ε for getting arbitrarily close to the optimal strategy asymptotically (the ε-decreasing strategy, with an ε function carefully chosen, achieves zero regret).

The lever with the highest estimated mean is always pulled except when a random lever is pulled instead with an $\varepsilon_t$ frequency where t is the index of the current round. The value of the decreasing $\varepsilon_t$ is given by $\varepsilon_t = \min\{1, \varepsilon_0/t\}$ where $\varepsilon_0 > 0$. The choice of $\varepsilon_0$ is left to the user. Beside ε-decreasing, Four other strategies are presented in (Auer et al., 2002).

Those strategies are not described here because the experiments done by (Auer al., 2002) seem to show that, with carefully chosen parameters, ε-decreasing is always as good as other strategies.

Compared to the standard multi-armed bandit problem with a fixed set of possible actions, we face a fast change in the contextual recommendation. Old documents may expire and new documents may emerge every day.

Therefore it may not be desirable to do the exploration all at once at the beginning as (Even-Dar et al., 2003) or monotonically decrease the effort on exploration as the decreasing strategy as (Mannor & Tsitsiklis, 2003).

### 2.2 Bandit Algorithms in the Contextual Recommender System Field

Few research works are dedicated to study the contextual bandit problem on contextual recommendation. In (Wei et al., 2010), the authors extend the ε-greedy strategy by updating the exploration value ε dynamically. At each iteration, they run a sampling procedure to select a new ε from a finite set of candidates.

The probabilities associated with the candidates are uniformly initialized and updated with the **Exponentiated Gradient (EG)** (Kivinen et al., 1997). This updating rule increases the probability of a candidate if it le documents to a user click. Compared to the ε-beginning and decreasing strategy, this technique improves the result.

However the technique takes a long time to find the optimal tradeoff and it does not consider the critical level of the user's situation on the exploration strategy.

In (Lihong et al., 2010), authors' model personalized recommendation of news documents as a contextual bandit problem, a principled approach in which a learning algorithm sequentially selects documents to serve users based on contextual information about the users and documents, while simultaneously adapting its document-selection strategy based on user-click feedback to maximize the total number of user clicks.

This work proposes a new, general contextual bandit algorithm that is computationally efficient; however they do not address the problem of exr/exp-tradeoff in their **LINUCB algorithm** for contextual bandit problem.

We summarize the related work in Table 1:

| ALGORITHM | E-DYNAMIC | CRITICAL LEVEL OF SITUATION | CONTEXTUAL RECOMMENDA-TION |
|---|---|---|---|
| E-BEGINNING | ✓ | x | x |
| E-DECREASING | ✓ | x | x |
| EG | ✓ | x | ✓ |
| LINUCB | x | x | ✓ |
| LINEARIZED -E-GREEDY() | ✓ | ✓ | ✓ |

Table 1: exr/exp-tradeoff algorithm

As it is shown by the Table 1, none of the mentioned works tackles both problems of making dynamic exr/exp and considering the user's situation on the exr/exp strategy. This is precisely what we intend to do with our approach, exploiting the following new features:

**Feature 1:** In (Wei et al., 2010) authors use a smart bandit algorithm to manage the exr/exp strategy, however they need a phase of training which may take time to converge to the optimal exr/exp-tradeoff. Inspired by the sampling theory, we propose for exr/exp-tradeoff a calibration, which consists in estimating the discrete probabilities of the rewards of the clicked and the non-clicked documents already recommended, then we build a linearized probability density distribution.

The exr/exp-tradeoff that maximizes a utility function, represented by both distributions, is then selected. We hope that this strategy will improve the global rewards of the system from the beginning of the recommendation.

**Feature 2:** Our intuition is that, the fact of considering the critical level of the situation on managing the exr/exp-tradeoff, will improve the long term rewards of the recommender system, which is not considered anywhere as far as we know.

## 3. Contextual Recommender System Model

In this section, we briefly describe our CRS.

In our CRS, documents' recommendation is modeled as a multi-armed bandit problem with context information of the user's situation, where the situation u is represented as a triple whose features X are the values assigned to each dimension: $u = (X_l, X_t, X_s)$, where $X_l$ (resp. $X_t$ and $X_s$) is the value of the location (resp. time and social) dimension. Suppose the user is associated to: the location "48.8925349, 2.2367939" from his phone's GPS; the time "Mon Oct 3 12:10:00 2011" from his phone's watch; and the meeting with Paul Gerard from his calendar.

To build the situation, we associate to this kind of low level data, directly acquired from mobile devices capabilities, more abstracted concepts using ontologies reasoning means, and then we get "The user is in a restaurant with a financial client, and it is a work day".

Formally, a bandit algorithm proceeds in discrete trials t = 1,…T. For each trial t, the algorithm performs the following tasks:

**Task 1:** Let $u_t$ be the current user's situation, and US={$u_1,...,u_n$} the set of past situations. The system compares $u_t$ with the situations in past cases in order to choose the most similar one $u_p$ using the following semantic similarity:

$$u_p = \arg\max_{S^i \in PS} \left( \sum_j \alpha_j \cdot sim_j\left(X_j^c, X_j^i\right) \right) \qquad (1)$$

In equation 1, $sim_j$ is the similarity metric related to dimension j between two situation vectors and $\alpha_j$ the weight associated to dimension j. $\alpha_j$ is not considered in the scope of this paper, taking a value of 1 for all dimensions.

The similarity between two concepts of a dimension j in an ontological semantic depends on how closely they are related in the corresponding ontology (location, time or social). We use the same similarity measure as (Wu & Palmer, 1994) defined by equation 2:

$$sim_j\left(X_j^t, X_j^c\right) = 2 * \frac{deph(LCS)}{(deph(X_j^c) + deph(X_j^t))} \qquad (2)$$

Here, LCS is the Least Common Subsumer of $X_j^c$ and $X_j^i$, and *depth* is the number of nodes in the path from the node to the ontology root. It observes the current user's situation $u_p$ and a set $A_t$ of arms together with their feature vectors $x_{t,a}$ for $a \in A_t$. The vector $x_{t,a}$ summarizes information of both user's situation $u_t$ and arm a, and is referred to as the context.

**Task 2:** Based on observed rewards in previous trials, it chooses an arm $a_t \in A_t$, and receives reward $r_{t,a_t}$ whose expectation depends on both the user's situation $u_t$ and the arm $a_t$.

**Task 3:** It improves its arm-selection strategy with the new observation, ($x_{t,a_t}, a_t, r_{t,a_t}$).

In tasks 1 to 3, the total T-trial reward of A is defined as $\Sigma_{t=1}^{T} r_{t,a_t}$ while the optimal expected T-trial reward is defined as $E\left[\Sigma_{t=1}^{T} r_{t,a_t^*}\right]$ where $a_t^*$ is the arm with maximum expected reward at trial t. Our goal is to design the bandit algorithm so that the expected total reward is maximized.

In the field of document recommendation, we may view documents as arms.

When a document is presented to the user and this one selects it by a click, a reward of 1 is incurred; otherwise, the reward is 0.

With this definition of reward, the expected reward of a document is precisely its Click Through Rate (CTR).

The CTR is the average number of clicks on a recommended document, computed diving the total number of clicks on it by the number of times it was recommended. Consequently, choosing a document with maximum CTR is equivalent, in our bandit algorithm, to maximizing the total expected rewards.

# 4. Exploration/Exploitation Tradeoff

In the rest of this section, we present the proposed exr/exp-tradeoff approaches, and then we show how integrate that in the contextual bandit algorithm.

## 4.1 Dynamic Tradeoff Adaptation

To define the best exploitation policy, the system should follow the document reward evolution and regulate the exr/exp-tradeoff. The regulation can be made by maximizing a utility function.

According to the sampling theory, the behavior of a random sample is the same for all the population, so the exr/exp-tradeoff allowing to maximize the utility function in a random sample of documents allows to maximize the same utility function in all the documents of the stream.

The approach we propose for exr/exp-tradeoff calibration consists in estimating the discrete probabilities of the rewards of the clicked and the non-clicked documents already recommended and then we build a linearized probability density distribution. The exploration that maximizes a utility function, represented by both distributions, is then selected.

### 4.1.1 REWARD DISTRIBUTION MODELING

The probability that a random document has a particular reward is equal to the number of documents having the same reward divided by the total number of documents:

$$P(X = reward) = \frac{|\{d|CTR(d) = reward\}|}{d} \quad (3)$$

As reward values are very distinct, they tend to be equiprobable ($\{|d|CTR(d) = reward|\} = 1$ or 0). Indeed, it is very difficult to find two or many documents having exactly the same reward. Consequently the reward probability distribution tends to be uniform.

Instead of computing the probability that a document has a certain reward, we compute the probability that the document reward belongs to an interval. We define a set of intervals enough reduced so that the document rewards belonging to the same interval are really closed. We define n adjacent intervals $I_1, I_2 \ldots I_n$ having the same ray as follows:

$$I_i = [reward_{i-1}, reward_i] \quad (4)$$

where: $reward_0 = \min_d CTR(d)$ and $reward_n = \max_d CTR(d)$.

The number of intervals is proportional to the sample size; indeed the great is the number of documents, the great is the definition field of the document rewards. We define n as the half of the total number of documents: n = d/2. The probability that a document reward X belongs to an interval is given by:

$$P(reward_i \leq X < reward_{i+1}) = \frac{|\{d|CTR(d) \in ]reward_i, reward_{i+1}]\}|}{d} \quad (5)$$

There are many methods to estimate the probability law parameters followed by the document rewards, for example the parametric regression and the maximum likelihood estimation method (Zhang et al., 2005) need to know the distribution law in advance. But, in an experimental context, many limits of these methods can be noted.

Indeed, even though the assumption that the rewards follow a known distribution density could be acceptable, but it strongly depends on the number of documents within the sample.

The sample size must be enough important to obtain non-asymmetric estimations.

To solve these problems, instead of assuming that the distributions are known, we build them by estimating the discrete probabilities of the rewards of the clicked and the non-clicked documents displayed at a certain time and then by linearizing these probabilities to obtain the corresponding probability density distributions.

A utility function, to be optimized, is then represented by using these distributions. The best exploration is the one that maximizes this utility.

### 4.1.2 PROBABILITY DISTRIBUTION LINEARIZATION

The linearization consists in dividing the domain of a function it into a set of intervals such as the representative curve of the restriction of that function in each interval can be assimilated to a linear curve.

We use this technique to linearize the representative curve of the probability density distribution of the rewards. We assume that this function exists and we try to linearize it. As we do not know this function, we propose to linearize it using the corresponding discrete probabilities.

The first step of this process is to identify a set of linear intervals. We define a linear interval between two rewards $[reward_x, reward_y]$ where $reward_x < reward_y$ and all the points formed by $(s_i, p_i)$ fit a straight line, $s_i$ being a reward in that interval and $p_i$ being the discrete probability of $s_i$ computed according to Formula 3.

The linearity of a set of points is measured by the least squares method (Zhang et al., 2005).

The least squares method requires that a straight line be fitted to a set of points so that the sum of the squares of the distance of the points to the fitted line is minimized.

In our work, the detection of a linear interval is incrementally done by considering all the points $(s_i, p_i)$, ordered in an increasing order of the rewards, as indicated in Algorithm 1.

This algorithm consists of adding a point to a given set of points representing a straight line and computing an error which measures the standard deviation between that "new" set and a linear curve. If this error is below a "threshold", the considered point is definitely added to that set, and the next point is then considered.

Otherwise this point is removed from the set, and we continue the search of a new linear interval, and so on until the last point of the distribution.

---

**Algorithm 1** *linearize()*

---

**Input:** *M*
**Output:** *P*
c = 1, P = ∅, threshold error = 0.0001
**For i = 0 to M do**
a)  P ← P ∪ {i};
b)  determine the equation of the line $L_c : y(x) = a + b*x$ based on the linear regression for all points $(s_j, p_j)$ ∀j ∈P;
c)  compute the standard deviation error between the points $(s_j, p_j)$ ∀j ∈ P and the line $L_c$:

$$E = \sum_{j \in P} d^2\left((s_j, p_j), L_c\right) = \sum_{j \in P} \left(\frac{a+b*s_j - p_j}{\sqrt{a^2+1}}\right)^2 \quad (6)$$

d)  **if** E > threshold error;
    i. //a class of points is formed
       $C_c = (d_c, f_c, a_c, b_c)$ where, $d_c = min(s_j)$, $f_c = max(s_j)$ ∀j ∈ P, $a_c$ and $b_c$ are the coecients of the equation of the line $y = a_c + b_c x$ derived using the linear regression of all the points $(s_j, p_j)$ where j ∈ P {i};
    ii. P ←{i}; // re-initialize P
    iii. c ← c + 1;
e)  **end if**
**End for**
*Normalize()*;

---

In Algorithm 1, c is the index of the classes of points with rewards within a linear interval, and M the total number of the points of the form $(s_i, p_i)$, ranked in an increasing order of their rewards.

After the searching linear intervals, the *Normalize* method (Alg.1) is executed so that all lines form a continuous representation. This is done through the following steps.

1. Transforms the curve representation into a continuous one by relying the extremities of two adjacent classes.

This liaison is done as follows: for two adjacent linear classes $C_c$ and $C_{c+1}$, rely $f_c$ and $d_{c+1}$ with a line having as equation $y = α_c + β_c x$. This line should pass through the points $(f_c, a_c + b_c f_c)$ and $(d_{c+1}, a_{c+1} + b_{c+1} d_{c+1})$, so:

$$α_c = \frac{a_c + b_c * f_c - a_{c+1} + b_{c+1} * d_{c+1}}{f_c - d_{c+1}}$$

(7)

$$β_c = a_c + b_c * f_c - \frac{a_c + b_c * f_c - a_{c+1} + b_{c+1} * d_{c+1}}{f_c - d_{c+1}}$$

(8)

2. Normalizes the coecients $a_c$, $b_c$, $α_c$ and $β_c$ such that:

$$\int_{reward_0}^{reward_m} f(x)dx = 1 \quad (9)$$

Equation 9 is the fundamental property of the probability density functions. As $\int_{reward_0}^{reward_m} f(x) dx$ is the surface formed by the graphical representation of f and X-coordinate axis, the coefficients $a_c$, $b_c$, $α_c$ and $β_c$ are divided by this surface. This surface becomes unit. This surface is computed as the sum of the surfaces formed by all the linear intervals and the X-coordinate axis.

The figure 1 illustrates a linearization performed on all clicked and not-clicks documents for our sampling.

It shows that the linearization probabilities reward tends to be an exponential for clicked documents and a Gaussian for not clicked documents.

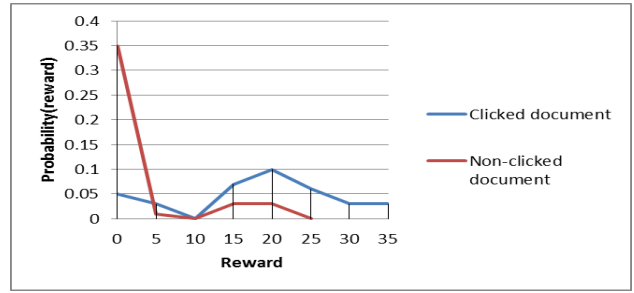

Figure 1. Linearization performed on all clicked and not-clicks documents

### 4.1.3 EXPLORATION OPTIMIZATION

In order to define the value of the exr/exp we have to determine a threshold *o* that maximizes the theoretical value of UF:

$$o = \arg\max_{o \in I_i}(UF) \quad (10)$$

The UF is the utility function which is generally written as follows:

$$UF = a * CD - b * NCD \quad (11)$$

In formula 11, a and b: positive constants; CD is the number of clicked documents selected and NCD is the number of non-clicked documents selected.

$$CD = EXP * TCD \qquad (12)$$

$$NCD = EXR * TNCD \qquad (13)$$

TCD and TNCD represent the total number of clicked and non-clicked examined documents.

Exr is the probability of making an exploration and in our case we suppose that it is equal to the probability that a document is selected when it was previously clicked and its denoted p(reward>o |r).

Exp is the probability of making an exploitation and in our case we suppose that it is equal to the probability that a document is selected when it was previously clicked and its denoted p(reward>o |s).

Where p(reward>o |s) (resp. p(reward>o |r)) represents the surface formed by the curve of function $f$ corresponding to the clicked (resp. non-clicked) documents and the X-coordinate axis, and p(r) = CD/N (resp. p(s) = NCD/N) is the probability that a document is clicked (resp. non-clicked).

Based on Bayes transformation rule, we obtain:

$$CD = \frac{p(reward \geq o|r) * p(reward > o)}{p(r)} * TCD \qquad (14)$$

$$NCD = \frac{p(reward \geq o|s) * p(reward > o)}{p(r)} * TNCD \qquad (15)$$

Utility done by equation 11 is equivalent to:

$$UF = p(reward>o) * N * (a * p(reward>o|r) + b * p(reward>o|s)) \qquad (16)$$

The retained threshold and in consequent exr/exp are the values that allows to maximize UF.

#### 4.1.4 Linearized -e-greedy

To improve exploitation of the ε-greedy algorithm, we propose to use the linearization of clicked and not-clicks documents and the optimization of the utility function at the beginning of the ε-greedy to define the exploration parameters ε.

The algorithm 2 summarizes the functional steps of *linearized-ε-greedy()* method.

---

**Algorithm 2** *linearized-ε-greedy()*

---

**Input:** *N, dc, ndc*
**Output:** *D*
**D**= ∅
// linearize probability distributions of these samples
*linearize(*dc*); linearize(ndc);*
//Compute the exploration exploitation that maximizes the utility function

$$o = \arg\max_{o \in I_i}(UF)$$

ε= p(reward>o |r)=
**For I =1 to N do**
    q = Random({0,1})
    $d_i = \begin{cases} argmax_{\,d \in (dc-di)}\,(CTR(d)) & \text{if } q \leq \varepsilon \\ Random(d) & \text{otherwise} \end{cases}$
    D = D ∪ $d_i$
**Endfor**

---

In Algorithm 2, d, D are respectively the documents that can be recommended and the recommended document to the user, N the number of documents recommended to the user and dc, ndc are respectively the clicked and non-clicked examined documents.

### 5. Conclusion and Future Work

In this paper, we study the problem of exploitation and exploration in contextual recommender systems and propose a novel approach that adaptively balances exr/exp regarding the user's situation. In order to evaluate the performance of the proposed algorithms, we plan to compare our algorithms with the other standard exr/exp strategies, using an industry leading performance based CRS with real online event data.